\documentstyle[prl,epsfig,multicol,aps]{revtex}

\begin{document}
\draft
\title{Anharmonic properties of double giant dipole resonance}
\author{
V. Yu. Ponomarev$^{1,2}$,
P. F. Bortignon$^{1}$, R. A. Broglia$^{1,3}$,
and V. V. Voronov$^{2}$
}
\address{$^{1}$Dipartimento di Fisica, Universit\`a di Milano and
INFN
Sezione di Milano,
Via Celoria 16,
I-20133 Milano, Italy}
\address{$^{2}$Bogoliubov Laboratory of Theoretical Physics,
Joint Institute for Nuclear Research,
Dubna, Moscow region, Russia}
\address{$^{3}$The Niels Bohr Institute, University of Copenhagen,
Blegdamsvej 17,2100 Copenhagen \O, Denmark.}
\date{\today }
\maketitle

\begin{abstract}
A systematic microscopic study of the anharmonic properties of the
double giant dipole resonance (DGDR) has been carried out, for the first
time, for nuclei with mass number $A$ spanning the whole mass table.
It is concluded that the corrections of the energy centroid of the
$J^{\pi} = 0^+$
and $2^+$ components of the DGDR
from its harmonic limit are negative, have  a value of the order of
few hundred keV and follow  an $A^{-1}$ dependence.
\end{abstract}


\begin{multicols}{2}

\narrowtext

Many-fermion systems, from metals in bulk to atomic nuclei, display
collective degrees of freedom known as plasmons in the language of
solid state physics \cite{Mah83} and as giant resonances within the
framework of nuclear physics \cite{Bor98}. A very successful
description of these modes is provided by the time-dependent
mean-field theory \cite{Bbr94}, in the variety of versions known
as quasi-boson
approximation, random phase approximation, linear response theory
\cite{Pin63,Pno66,Lan64}, time-dependent Hartree-Fock theory \cite{Rsc80},
time-dependent local density approximation \cite{Msu90}, etc. At the
basis of all these, to a large extent, equivalent  theoretical
descriptions of the sloshing back and forth of electrons against ions,
of protons against neutrons, etc, one finds the small amplitude
approximation which identifies these modes with the one-phonon state
of an harmonic oscillator.
While it is true that a couple of fermions (particle-hole excitation)
has integer spin, its behavior cannot be identified for all relative
energies and momenta with that of a real boson, the range of validity
of this identification depending on the correlation energy of the pair.
On the
other hand, close to the ground state, fermion particle-hole
excitations  do behave as (quasi)bosons. In fact, the terms which, in
the equations of motion, are related to the non-bosonic contributions
of the commutation relations of pairs of fermions have random phases
leading to cancellations which reduce conspicuously the contribution
of the corresponding terms, eventually justifying the harmonic
approximation \cite{Pin63,Pno66,Lan64,Bop53}.
In any case, all degrees of freedom of a many-fermion system are
exhausted  by the degrees of freedom of the particles. Consequently,
although collective vibrations display small overlaps with each of the
(particle-hole) components of the wave function describing the mode, a
certain amount of over-counting is unavoidable.

With the advent of high fluency lasers and of high luminosity heavy
ion beams at relativistic energies, it is now possible to study
multi-plasmon states in bulk matter and in clusters \cite{Hab99},
as well as states of multiple excited giant dipole resonances in
atomic nuclei \cite{Aum98}, and thus test the limits of validity of
the harmonic paradigm in many-fermion systems \cite{Not1}.
In particular, the discovery of the double giant dipole resonance in
nuclei (DGDR) \cite{Aum98,Eml93,Mor94,Cho95,Bert99} and the
observation of small deviations from the harmonic picture concerning
the excitation energy and the spreading width, combined with the large
(up to a factor 2-3 enhancement) deviations of the associated Coulomb
excitation cross sections measured in relativistic heavy ion
collisions \cite{Aum98}, call for a better understanding of the role
anharmonicities play in the spectrum of the DGDR. In fact,
anharmonicities influence electromagnetic DGDR cross sections in
several ways:  a) the energy shifts of the DGDR states from the
harmonic values can affect in an important way the electromagnetic
cross section, in keeping with the exponential dependence of these
quantities with the Q-value of the process \cite{Win79}, b)
anharmonicities lead to changes in the E1 transition matrix elements
to preserve the energy weighted sum rule
(EWSR) \cite{Kur96} which eventually reinforce these effects,
c) anharmonicities which are a consequence of the mixing of states
with different number of phonons, give rise to many paths, other than
the (harmonic) two-step one, to excite the DGDR in electromagnetic
processes.
While all these questions inspired much theoretical work,
\cite{Cat89,Pon94,Nis95,Vol95,%
Pon96,Pon97,Bor97,Ber97,Lan97,%
Ham99,Ber99}, no clear picture has emerged of the DGDR anharmonicity
question, let alone an explanation of the ``Coulomb excitation anomaly".
In particular, no consensus exists concerning the mass-number
dependence of the energy shifts from the harmonic values.

In this Letter we present the results of the first,
systematic calculation of the
spectrum of the DGDR, carried out in a complete one- and two-phonon
basis (the effect of the 3 phonon states on the anharmonicity being
arguably small \cite{Pon96}) for nuclei with mass number $A$
spanning the whole mass table. It will be concluded that the energy
shift (lowering) of the energy centroid  of the
$J^{\pi} = 0^+$ and $2^+$ components of the DGDR from the harmonic
limit is rather modest (few hundreds of keV) and display a clear
$A^{-1}$ dependence. The solution of the Coulomb excitation anomaly is
thus likely to be found elsewhere \cite{Not2}.

The Hamiltonian used in describing the system contains, aside from a
mean field term which determines the single-particle motion of protons
and neutrons, a monopole pairing interaction and a separable
multipole-multipole force with strengths adjusted so as to reproduce
the odd-even mass differences and the spectrum of low-lying vibrations
and of giant resonances respectively \cite{Bor98,Rsc80,Boh75,Sol92}.
In particular, the strength of the isovector dipole-dipole term was
fixed by fitting the observed energy centroid of the GDR in
each nucleus or, lacking this information, the value emerging from the
energy systematics ($80 \cdot A^{-1/3}$ MeV).

The basis of one-phonon states was obtained by diagonalizing the
Hamiltonian  in the quasiparticle random phase approximation. The
basis of two-phonon states was constructed by coupling two one-phonon
states to total angular momentum and parity $J^{\pi} = 0^+$ and $2^+$,
in keeping with the quantum numbers of the DGDR states. 
The two-phonon basis thus include, aside from the states
$[1^-_i \times 1^-_{i'}]_{0^+(2^+)}$, where the
subindex $i$ is used to distinguish between the different one-phonon
dipole states arising from the shell structure of the system, also
two-phonon states made up of $0^+$, $1^-$, $2^+$, $3^-$, and $4^+$
phonons.
All one-phonon states, up to
40-50~MeV of excitation energy and contributing with more than 1.0\%
to the EWSR (0.2\% in the case of dipole modes)
have been included in the calculations.
This choice leads, for the  $J^{\pi} = 2^+$ component of the DGDR in
heavy nuclei, to a two-phonon basis containing of the order of 10$^3$
states.

\begin{figure}[htb]
\begin{center}
\epsfig{figure=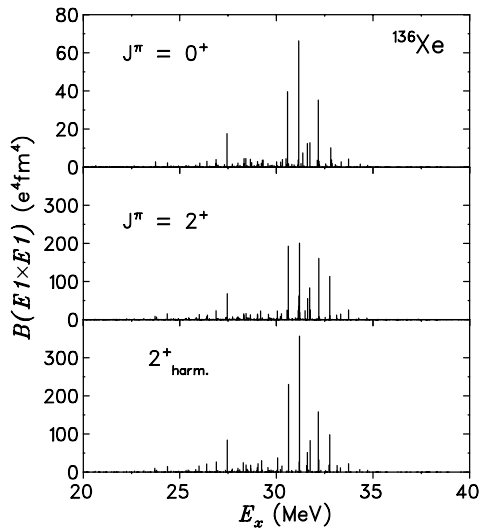,width=6.2cm,angle=0}
\end{center}
\caption{Energy distributions of the $B(E1 \times E1)$ values
associated with the excitation of the $0^+$ and $2^+$ components of the
DGDR in $^{136}$Xe,
in comparison with the same quantity for the $2^+$ component in the
harmonic limit. Scales are chosen proportionally to $(2J+1)$.
\label{xe} }
\end{figure}

The Hamiltonian written in terms of quasiparticles and phonons
\cite{Sol92}
is diagonal in the space of one- and two-phonon states separately,
but contain terms coupling one- to two-phonon states.
Diagonalizing this Hamiltonian, we obtain the total wave functions
$\Psi^{\nu}_{J}$  and the corresponding eigenvalues  from which the
results displayed in Table~\ref{tab1} have been obtained.
In the second column of this table, the percentage of the
Thomas-Reiche-Kuhn
EWSR exhausted by the selected one-phonon dipole states is displayed,
while the third column contains the percentage of the EWSR for the
DGDR (calculated in Ref.~\cite{Kur96})  exhausted relatively to the sum
of the $0^+$ and $2^+$ components.
The small differences observed between the percentage of the EWSR
exhausted by the DGDR and the GDR is mainly due to the fact that the
ground state is
considered in the calculations as the one-phonon vacuum,
the ground state
correlations  arising from the interaction between multi-phonon
configurations not being taken into account.
In the columns four and five the energy shifts
$\Delta E_c (J^{\pi})$ of the centroids of the $J^{\pi} =
0^+$ and $2^+$ members of the DGDR  with respect to the harmonic
predictions are reported.
In Fig.~\ref{xe}
we show the quantity
\[
B_{\nu}([E1 \times E1]_{J}) = \left | \sum_{i}
 \langle \Psi^{\nu}_{J}
| E1 | \Psi^{i}_{1^-} \rangle  \cdot
 \langle \Psi^{i}_{1^-} | E1 | \Psi_{g.s.} \rangle \right |^2
\]
for the different (two-phonon) states $\nu$, eigenstates of the total
Hamiltonian with angular momentum and parity $0^+$ and $2^+$ of
the nucleus $^{136}$Xe.
The same calculations have been repeated in a basis containing a
single two-phonon state $[1_{i_0}^- \times 1_{i_0}^-]_{0^+(2^+)}$, where
$1^-_{i_0}$ is the GDR mode carrying the largest fraction of the EWSR,
and all one-phonon states so as to reproduces as far as possible the
harmonic scenario within the framework of the present microscopic
calculation. We shall discuss these results before discussing those
of the full calculation.

The diagonalization in the reduced space leads to a breaking of the
$[1_{i_0}^- \times 1_{i_0}^-]_{0^+(2^+)}$, and thus to a set of states
with $J^{\pi} = 0^+$ and $2^+$,
one of which carries about 95\% of the two-phonon configuration
oscillator strength.
The energy shift of this state from the
energy of the (non-interacting) two-phonon configuration is shown in
Table~\ref{tab2} in the columns labeled ``Sum".
There are two mechanisms contributing to this shift:
the first one is associated with the Pauli principle corrections.
Excluding four-quasiparticle
configurations which violate Pauli principle reduces somehow the
collectivity of
two-phonon configurations. One thus expects
a downward shift for isovector phonons like e.g. the DGDR
(cf. column I of Table~\ref{tab2} displaying the results
obtained including only Pauli principle like-processes).
This shift  is found to scale  with $A^{-1}$ as expected from general
arguments \cite{Not4} and
simple models ~\cite{Boh75,Ham99,Ber99,Bes74}.
The second mechanism arise from the interaction
of the $[1^-_{i_0} \times 1^-_{i_0}]_{J^{\pi}}$ configuration
with all one-phonon states.
The energy shifts arising from this interaction are given in columns II
of Table~\ref{tab2}.
A strong cancellation with the first contribution is found,  although
not as complete as that reported  in
ref.~\cite{Ham99}, where estimates of the two contributions to the total
energy shift under discussion  have been carried
out within a schematic model.
This (second) contribution arising from the interaction of two-phonon
configurations with one-phonon states is found, in the present
simplified calculations, not to have any simple dependence  with $A$,
as it arises from the coupling of the single two-phonon configuration
chosen, to relatively few one-phonon configurations lying close in
energy and displaying a moderate value of the coupling matrix elements.

 Carrying the diagonalization in the full
 two- and one-phonon space, the contribution to the DGDR energy centroid 
 associated with the second mechanism
 vanishes because the trace of the matrices is conserved independently
 on the values of non-diagonal matrix elements.
 Consequently, the centroid of each of the DGDR configurations 
 (diagonal elements) remain
 at the non-interactive value,
 eventually modified by the Pauli principle for two-phonon corrections.
The corresponding energy shifts $\Delta E_c (J^{\pi})$
for the double magic nuclei $^{40}$Ca and $^{208}$Pb obtained in the
present calculation (cf. Table~\ref{tab1} columns  four and five) are
of the same order of magnitude as those obtained in microscopic
calculations
in Refs.~\cite{Cat89,Nis95,Lan97}.
On the other hand, the mixing between one- and two- phonon states
obtained for the other nuclei (cf. Table~\ref{tab2}) are larger than
for $^{40}$Ca and $^{208}$Pb. This is in keeping with the fact that
doubly-magic nuclei are much more rigid than the semi-magic ones.
The most collective
$[1_i^- \times 1_i^-]_{0^+(2^+)}$ configuration
in general prefers to mix
with either [LEOR$\times$HEOR] or other
$[1_i^- \times 1_{i'}^-]_{0^+(2^+)}$
configurations where L(H)EOR is the low (high) energy octupole
resonance.

\begin{figure}[htb]
\begin{center}
\epsfig{figure=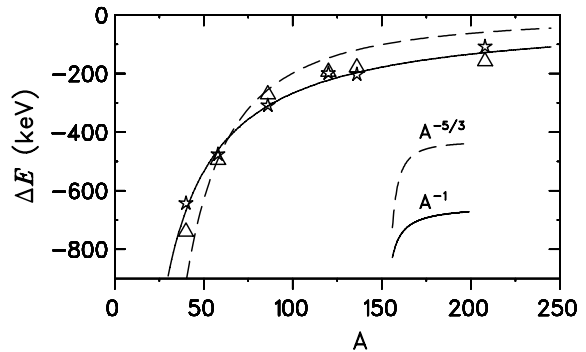,width=7.5cm,angle=0}
\end{center}
\caption{Shift of the DGDR energy centroid
($0^+$ - stars  and $2^+$ triangles)
from the harmonic limit. The continuous and dashed  curves represent
fits, assuming an A$^{-1}$ and an A$^{-5/3}$ dependence respectively,
of the results of the microscopic calculations.
\label{shift} }
\end{figure}

 From the systematic calculations one can extract the $A$-dependence of
the energy shifts  of the centroids from the harmonic limits.
For this purpose, the value of the energy shifts of the DGDR are shown
in Fig.~\ref{shift} as a function of $A$.
The continuous curve represents an $A^{-1}$ fitting to the data, while
the dashed line shows the $A^{-5/3}$ dependence
obtained in the variational time-dependent
approach of Ref.~\cite{Ber97}.
The results of our calculations follow quite accurately the $A^{-1}$
behavior, even if
both doubly- and semi-magic nuclei have been included in the
systematics.
Weighting equally the $0^+$ and $2^+$ components of the DGDR we obtain
from a $\chi^2$ analysis of the results displayed in Fig.~\ref{shift},
$\Delta E= b \cdot A^{-\alpha}$  with $\alpha = 1.08 \pm 0.06$ and
$b = -37 \pm 8$~MeV.

Although our conclusion on the $A^{-1}$ dependence
is based on calculations within a specific model,
general arguments \cite{Not4} and estimates \cite{Boh75,Ham99,Ber99,Bes74}
support it. Different $A$-dependence of the energy shift are discussed 
in \cite{Ber99}, in terms of the number of active nucleons. 
The present results indicate that in the case of the GDR this number is 
indeed of the order of $A^{2/3}$, as for the $\Omega$ factor in \cite{Not4}.

We conclude  that the deviation of the energy centroid
of the double giant dipole resonance  from the harmonic limit displays
a behavior with mass number $A$ typical of that associated with the
global properties characterizing the system, like e.g. the energy
centroid of the giant dipole resonance.

Discussions with G. F. Bertsch, K. Hagino, I. Hama\-moto, and B.
Mottelson are gratefully acknowledged.
We thank J. Bryssinck for help.
V. Yu. P. acknowledges a support from INFN and NATO.

\begin{table}[htb]
\caption{Percentage of the EWSR exhausted by
the GDR and DGDR of the atomic nuclei indicated in the first column.
In column 4 and 5 is displayed the
anharmonicity shift $\Delta E_c (J^{\pi})$
of the energy centroid of the $J^{\pi} = 0^+$
and $2^+$ components of the DGDR
from its harmonic limit.
}
\label{tab1}
\begin{tabular}{ccccc}
A     & \multicolumn{2}{c} {EWSR, \%} & \multicolumn{2}{c}
{$\Delta E_c (J^{\pi})$, keV}\\
Nucl. & GDR & DGDR & $J^{\pi}=0^+$ & $J^{\pi}=2^+$ \\
\hline
$^{40}$Ca  & 104. & 103. & -643 & -740 \\
$^{58}$Ni  & 104. & 103. & -476 & -495 \\
$^{86}$Kr  & 106. & 105. & -309 & -271 \\
$^{120}$Sn & 106. & 105. & -199 & -194 \\
$^{136}$Xe & 103. & 102. & -203 & -179 \\
$^{208}$Pb &  94. &  94. & -108 & -158 
\end{tabular}
\end{table}

\begin{table}[htb]
\caption{Energy shift  of the two states
$[1^-_{i_0} \times 1^-_{i_0}]_{J^{\pi}}$
($J^{\pi} = 0^+$ and $2^+$) with respect
to the harmonic value $2\hbar\omega({1_{i_0}^-})$.
The label $i_0$ indicates the component
of the GDR carrying the largest fraction of the EWSR.  The calculations
have been carried out in a basis which includes only
the two-phonon configuration $[1^-_{i_0} \times 1^-_{i_0}]_{J^{\pi}}$
and a complete set of
$0^+$ ($2^+$)  one-phonon states.
The contributions to the energy shift arising
from Pauli principle corrections and due to the interaction of the
two-phonon configuration
with one-phonon configurations are shown separately in I and II,
respectively.
}
\label{tab2}
\begin{tabular}{crrrrrr}
A     & \multicolumn{3}{c} {$0^+$} & \multicolumn{3}{c} {$2^+$}
\\
Nucl. & I & II & Sum & I & II & Sum \\
\hline
$^{40}$Ca   & -577  &  +274  &  -302 &  -740 &   +534  &  -206 \\
$^{58}$Ni   & -387  &  +667  &  +280 &  -486 &   +507  &  +21  \\
$^{86}$Kr   & -240  &  +103  &  -137 &  -291 &   +227  &  -64  \\
$^{120}$Sn  & -163  &  +181  &  +18  &  -223 &   +204  &  -19  \\
$^{136}$Xe  & -142  &  +87   &  -55  &  -186 &   +171  &  -15  \\
$^{208}$Pb  & -104  &  +129  &  +24  &  -137 &   +93   &  -44  \\
\end{tabular}
\end{table}

\end{multicols}


\begin{references}
\bibitem{Not1}
It can be argued that much work concerning this subject have been
carried out (cf. e.g. Refs. \cite{And84} and \cite{Enz92}), in
particular in connection with the study of the low-lying vibrations of
nuclei \cite{Boh75}. While this is true, the comparison between
theories developed to deal with the associated violations of Pauli
principle, like e.g. Nuclear Field Theory (cf. e.g. \cite{Bor77} and
Refs. therein) or Quasiparticle-Phonon Model \cite{Sol92}, and
experiment has been, in many cases, blurred by
the (relatively) small phase space available to the nucleons in the
low-lying vibrations, and the role specific single-particle orbits
play in determining the properties of the modes.
\bibitem{Not2}
As a matter of fact and strictly speaking, such a conclusion must wait
the explicit calculation of the ratio of the Coulomb excitation cross
section associated with the (mainly) two-phonon and (mainly) one-phonon
states. In fact, to avoid the complications of combining nuclear
structure with nuclear reaction theories to eventually carry out a
comparison
with the experimental data, and be able to extract, in the simplest way,
the $A$-dependence of the two-phonon anharmonicities, we have avoided
explicit reference to the one-phonon states.
In keeping with the fact that the coupling  to one-phonon states leads
only to a breaking of the strength (damping width), but
not to a change in the energy centroid, we are able to concentrate in
the central issue which is at the basis of the DGDR energy centroid
shift from the harmonic value, namely the modification of the phonon
structure due to Pauli principle corrections.
\bibitem{Not4}
The lowest order, Pauli principle correction correspond to fourth order
(``butterfly") diagrams (cf. e.g. Fig. 9 of Ref. \cite{Bes76}) leading
to the energy shift which can be quantitatively written as
$\Delta E^{(4)}_{Pauli}\sim \Lambda^4\Omega/Den^3$. In this expression
$\Lambda\sim (\hbar\omega/2C)^{1/2}$ is the particle-vibration coupling
strength, $\hbar\omega\sim A^{-1/3}$ being the energy of the mode
(resonance) to which the particle couples, $C\sim A$ the associated
restoring force. Consequently $\Lambda\sim A^{-2/3}$. The energy
denominators arise from energy differences between giant resonances and
quasiparticle pairs  and display the typical $A$-dependence of the GDR
energy, thus $Den\sim A^{-1/3}$. The quantity $\Omega\sim A^{2/3}$ (cf.
e.g. \cite{Bes69}) measures the degeneracy of the phase space where
fermions are allowed  to correlate to produce the collective GDR
(closed loop summations in field theoretical terms). Consequently
$\Delta E^{(4)}_{Pauli}\sim A^{-1}$.
\end{references}
\end{document}